\documentclass[11pt, a4paper]{article}

\usepackage[margin=2.5cm]{geometry}
\usepackage{amsmath, amssymb, amsthm}
\usepackage{graphicx}
\usepackage{xcolor}
\usepackage{booktabs}
\usepackage{hyperref}
\usepackage{enumitem}
\usepackage{tcolorbox}
\usepackage{tikz}
\usetikzlibrary{arrows.meta, positioning, shapes.geometric}
\usepackage{listings}
\usepackage{fancyhdr}
\usepackage{titlesec}
\usepackage{parskip}
\usepackage{mdframed}
\usepackage{caption}
\usepackage{microtype}
\usepackage{mathpazo}   
\usepackage[round, authoryear]{natbib}
\hypersetup{colorlinks=true, citecolor=epidblue, linkcolor=epidblue, urlcolor=epidblue}
\definecolor{epidblue}{RGB}{31, 97, 141}
\definecolor{epidgreen}{RGB}{30, 132, 73}
\definecolor{epidred}{RGB}{185, 36, 28}
\definecolor{lightgray}{RGB}{245, 245, 245}
\definecolor{boxblue}{RGB}{214, 234, 248}
\definecolor{boxgreen}{RGB}{212, 239, 223}
\definecolor{boxyellow}{RGB}{253, 243, 207}
\definecolor{codegray}{RGB}{248, 248, 248}

\titleformat{\section}{\large\bfseries\color{epidblue}}{}{0em}{\thesection\quad}[\vspace{-0.3em}\rule{\linewidth}{0.8pt}]
\titleformat{\subsection}{\normalsize\bfseries\color{epidblue!80!black}}{\thesubsection}{0.8em}{}

\tcbuselibrary{skins, breakable}

\newtcolorbox{keybox}[1]{
  colback=boxblue, colframe=epidblue, fonttitle=\bfseries,
  title=#1, rounded corners, left=6pt, right=6pt, top=4pt, bottom=4pt,
  breakable
}
\newtcolorbox{intuitionbox}[1]{
  colback=boxgreen, colframe=epidgreen!80!black, fonttitle=\bfseries,
  title=#1, rounded corners, left=6pt, right=6pt, top=4pt, bottom=4pt,
  breakable
}
\newtcolorbox{warningbox}[1]{
  colback=boxyellow, colframe=orange!80!black, fonttitle=\bfseries,
  title=#1, rounded corners, left=6pt, right=6pt, top=4pt, bottom=4pt,
  breakable
}

\begin{document}

\begin{titlepage}
  \centering
  \vspace*{2cm}
  {\Huge\bfseries\color{epidblue} Bayesian Inference in Epidemic Modelling:\\[0.4em]
 \large\color{epidblue!70!black} A Beginner's Guide}\\[2em]

  \rule{0.4\textwidth}{1pt}\\[3.5em]

  {\large Illustrated with the SIR Model}\\[2em]
  {\small\textit{Authored by}}\\[1.5em]
  {\large\bfseries Augustine Okolie}\\[0.2em]
  {\small PhD Mathematics}\\[6em]

  \begin{tcolorbox}[width=0.78\textwidth, colback=boxblue, colframe=epidblue,
    rounded corners, halign=center]
    \textit{``All models are wrong but some are useful. Bayesian inference tells us \textbf{how} useful and \textbf{how wrong}.''}\\[0.3em]
    {\small George E.P.\ Box}
  \end{tcolorbox}

  \vfill
    {\small Lecture Notes $\cdot$ Mathematical Epidemiology $\cdot$ Bayesian Statistics}
\end{titlepage}

\tableofcontents
\newpage

\section{Why Do We Need Bayesian Inference?}

When we build an epidemiological model, say, for COVID-19, influenza or Ebola, we face a fundamental challenge: \textbf{we do not know the model's parameters $\theta$}. How infectious is the pathogen? How quickly do people recover? These are the numbers that drive the model, yet they must be inferred from noisy, unobserved or incomplete data.

There are two broad philosophies for doing this.

\medskip
\begin{tabular}{p{0.45\textwidth} p{0.45\textwidth}}
\toprule
\textbf{Frequentist approach} & \textbf{Bayesian approach} \\
\midrule
Find the single ``best'' parameter value (e.g.\ maximum likelihood). & Find the full \emph{distribution} of plausible parameter values. \\[4pt]
Reports a point estimate, perhaps with a confidence interval. & Reports a \emph{posterior distribution}, a curve showing how credible each parameter value is. \\[4pt]
Cannot easily incorporate prior knowledge. & Formally combines prior knowledge with new data. \\[4pt]
Asks: ``What is $\hat{\theta}$?'' & Asks: ``What is $P(\theta \mid \text{data})$?'' \\
\bottomrule
\end{tabular}

\bigskip
\begin{intuitionbox}{The Key Intuition}
Imagine you are trying to guess the weight of a person you have never met. Before seeing any data, you already know something: people rarely weigh 5 kg or 500 kg. That is your \textbf{prior}. Then you observe their height and build. That is your \textbf{data} (likelihood). Bayesian inference combines both into a \textbf{posterior}, your updated, data-informed belief about their weight.
\end{intuitionbox}

\section{The SIR Model}

The Susceptible-Infected-Recovered (SIR) model is the workhorse of compartmental epidemiology. It divides a population of size $N$ into three groups:

\begin{center}
\begin{tikzpicture}[node distance=2.8cm, >=Stealth, thick]
  \node[draw, rounded corners, fill=epidblue!15, minimum width=1.8cm, minimum height=1cm] (S) {\textbf{S}usceptible};
  \node[draw, rounded corners, fill=epidred!15,  minimum width=1.8cm, minimum height=1cm, right=of S] (I) {\textbf{I}nfectious};
  \node[draw, rounded corners, fill=epidgreen!15,minimum width=1.8cm, minimum height=1cm, right=of I] (R) {\textbf{R}ecovered};
  \draw[->] (S) -- node[above]{\small$\beta S I / N$} (I);
  \draw[->] (I) -- node[above]{\small$\gamma I$} (R);
\label{tikz:sir_flow}
\end{tikzpicture}
\end{center}

The dynamics are governed by a system of ordinary differential equations (ODEs):

\begin{align}
  \frac{dS}{dt} &= -\frac{\beta S I}{N}, \label{eq:dS}\\[4pt]
  \frac{dI}{dt} &=  \frac{\beta S I}{N} - \gamma I, \label{eq:dI}\\[4pt]
  \frac{dR}{dt} &=  \gamma I.\label{eq:dR}
\end{align}

\begin{keybox}{Parameters}
\begin{itemize}[nosep]
  \item $\beta$ is the \textbf{transmission rate}: how quickly susceptibles become infected per contact.
  \item $\gamma$ is the \textbf{recovery rate}: the fraction of infected individuals who recover per unit time.
  \item $\mathcal{R}_0 = \beta/\gamma$ \quad is the \textbf{basic reproduction number}: average new infections caused by one case in a fully susceptible population. If $\mathcal{R}_0 > 1$, the epidemic grows.
\end{itemize}
\end{keybox}

Each compartment represents a distinct epidemiological state that every individual
occupies at any given time:

\begin{itemize}
  \item \textbf{Susceptible} $S(t)$: individuals who have never been infected at time $t$ and
  have no immunity. They are at risk of contracting the disease upon contact with
  an infectious person. At the start of an outbreak, nearly the entire population
  sits in this compartment.

  \item \textbf{Infectious} $I(t)$: individuals who are currently infected at time $t$ and
  capable of transmitting the pathogen to susceptible individuals. The size of
  this compartment at any moment determines the force of infection driving the
  epidemic. In the basic SIR model, we assume that all infectious individuals
  are equally and immediately infectious, no latency period.

  \item \textbf{Recovered} $R(t)$: individuals who have cleared the infection at time $t$ and
  are assumed to have acquired \emph{permanent immunity}. They can neither be
  reinfected nor transmit the disease. In some diseases this is a reasonable
  approximation (e.g.\ measles); in others (e.g.\ influenza, COVID-19) immunity
  wanes over time, requiring extended models such as SIRS or SEIRS.
\end{itemize}

The arrows in the schematic representation of the standard SIR model (Figure \ref{tikz:sir_flow}) show the only permitted transitions: $S \to I$ (infection)
and $I \to R$ (recovery). There is no return path from $R$ to $S$, and no direct
route from $S$ to $R$, every individual must pass through the infectious state.
The rates labelling the arrows, $\beta SI/N$ and $\gamma I$, are precisely the
terms that appear in the ODEs \ref{eq:dS} -- \ref{eq:dR}.

\subsection{Solving the ODE: From Equations to Epidemic Curves}

The system of equations \eqref{eq:dS}, \eqref{eq:dR} has no closed-form analytical solution. Instead, we use
\textbf{numerical integration}: starting from initial conditions, we approximate the solution
by taking many small time steps, updating each compartment at every step.

\subsubsection*{The Euler idea (intuition)}

The simplest intuition comes from Euler's method. If we know the state at time $t$,
we can approximate the state a small step $\Delta t$ later by multiplying the current
rate of change by the step size and adding it to the current value

\begin{align}
  S(t + \Delta t) &\approx S(t) + \frac{dS}{dt}\,\Delta t
  &=&\; S(t) - \frac{\beta\, S(t)\, I(t)}{N}\,\Delta t, \\[6pt]
  I(t + \Delta t) &\approx I(t) + \frac{dI}{dt}\,\Delta t
  &=&\; I(t) + \left(\frac{\beta\, S(t)\, I(t)}{N} - \gamma\, I(t)\right)\Delta t, \\[6pt]
  R(t + \Delta t) &\approx R(t) + \frac{dR}{dt}\,\Delta t
  &=&\; R(t) + \gamma\, I(t)\,\Delta t.
\end{align}

Notice the internal consistency: whatever \emph{leaves} $S$ in the first equation
immediately \emph{enters} $I$ in the second, and whatever \emph{leaves} $I$ via
recovery immediately \emph{enters} $R$ in the third. The total population
$S + I + R = N$ is therefore conserved at every step, no one is created or destroyed,
only moved between compartments.

The step size $\Delta t$ controls the trade-off between speed and accuracy.
A large $\Delta t$ is fast but accumulates errors; a small $\Delta t$ is accurate
but slow. In practice we use more sophisticated solvers such as
\textbf{Runge-Kutta 4 (RK4)}, which evaluates the slope at four intermediate
points within each step and takes a weighted average.



The idea is identical to Euler,
\emph{follow the slope, step by step}, but with a much more accurate
estimate of which slope to follow. The \texttt{scipy} function \texttt{odeint} in Python uses an adaptive version of this internally, automatically shrinking $\Delta t$
in regions where the solution changes rapidly (e.g.\ near the epidemic peak)
and enlarging it where the solution is flat.



\subsection{Interpreting the SIR Dynamics}

\subsubsection*{Numerical solution and epidemic curves}

\begin{figure}[h!]
  \centering
  \includegraphics[width=\textwidth]{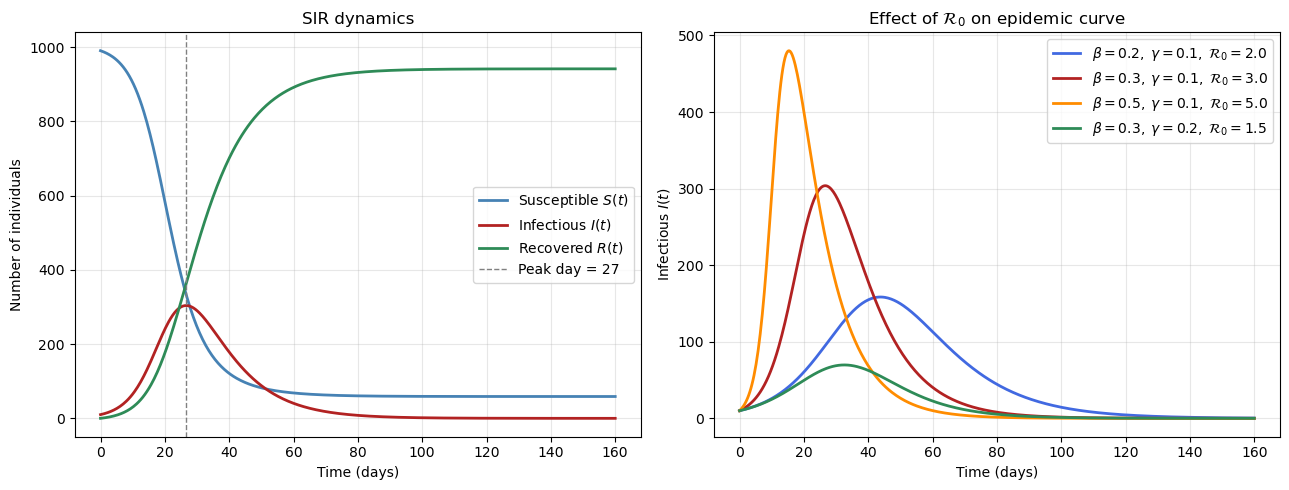}
  \caption{Numerical solution of the SIR model. $N=1000$, $I(0)=10$, $S(0)=990$, $R(0)=0$, with $\beta=0.3$, $\gamma=0.1$, $\mathcal{R}_0=3.0$.}
  \label{fig:sir_ode}
\end{figure}

As shown in Figure~\ref{fig:sir_ode}, the numerical solution reveals three
compartments evolving over time and the sensitivity of the epidemic curve to
the underlying parameters.

\medskip
\textbf{Left panel: the three compartments}. Three features stand out. First, $S(t)$ declines monotonically, once infected,
people do not return to the susceptible pool. Second, $I(t)$ rises, peaks, then
falls: the peak occurs precisely when
\[
  S(t^*) = \frac{N\gamma}{\beta},
\]
i.e.\ when there are no longer enough susceptibles to sustain exponential growth.
At this point the rate of new infections exactly equals the rate of recoveries,
and the infectious curve turns over. Third, $R(t)$ rises sigmoidally and plateaus
at the \emph{final epidemic size}, the total fraction of the population ever
infected \citep{Kermack1927}. Importantly, this plateau is strictly below $N$:
some susceptibles always escape infection, protected by the declining pool of
infectious individuals near the end of the outbreak.

\medskip
\textbf{Right panel: the role of $\mathcal{R}_0$ and the $\beta$-$\gamma$ interplay.} Varying $\beta$ and $\gamma$ while tracking $I(t)$ reveals that $\mathcal{R}_0$
is not controlled by either parameter alone, but by their \emph{ratio}.
Recall that $1/\gamma$ is the mean infectious period (in days). During that window,
an infected individual transmits at rate $\beta$, so:

\[
  \mathcal{R}_0 \;=\; \beta \times \frac{1}{\gamma}
  \;=\; \frac{\text{transmission rate}}{\text{recovery rate}}
\]

This creates a natural tension between the two parameters (See Table \ref{tab:example-disease}):

\begin{itemize}
  \item A \textbf{highly contagious but short-lived} infection (large $\beta$,
  large $\gamma$) can have a \emph{low} $\mathcal{R}_0$, the host recovers
  before spreading widely.

  \item A \textbf{moderately contagious but prolonged} infection (moderate $\beta$,
  small $\gamma$) can have a \emph{high} $\mathcal{R}_0$, the long infectious
  window compensates for slower per-contact transmission.

  \item Only when $\beta$ \emph{dominates} $\gamma$, i.e.\ transmission outpaces
  recovery, does $\mathcal{R}_0 > 1$ and the epidemic grow.
\end{itemize}

Higher $\mathcal{R}_0$ produces a \textbf{taller}, \textbf{earlier} epidemic peak
and a larger final size \citep{Keeling2008}. Crucially, two parameter pairs with the
\emph{same} $\mathcal{R}_0$ but different $\beta$ and $\gamma$ produce different
curve shapes: a higher $\gamma$ compresses the infectious period and sharpens the
peak even when $\mathcal{R}_0$ is unchanged. $\mathcal{R}_0$ alone therefore does
not fully characterise epidemic dynamics, the individual rates matter too.

\begin{table}[]
    \centering
    \begin{tabular}{lccccl}
    \toprule
    \textbf{Disease} & $\beta$ & $\gamma$ & $1/\gamma$ (days) & $\mathcal{R}_0$ & \textbf{Reference} \\
    \midrule
    Our SIR example  & 0.30 & 0.10 & 10 & 3.0          & --- \\
    Seasonal flu     & 0.50 & 0.33 & 3  & 1.5          & \citet{Biggerstaff2014} \\
    COVID-19 (early) & 0.25 & 0.10 & 10 & 2.5          & \citet{Liu2020} \\
    Measles          & 1.50 & 0.07 & 14 & $\approx 15$ & \citet{Guerra2017} \\
    Ebola            & 0.20 & 0.10 & 10 & $\approx 2$  & \citet{Althaus2014} \\
    \bottomrule
    \end{tabular}
    \caption{$\mathcal{R}_0$ estimates across diseases. Notice that measles achieves its explosive $\mathcal{R}_0 \approx 15$ through a
    combination of very high $\beta$ \emph{and} a long infectious period. Ebola, despite
    being lethal, has a modest $\mathcal{R}_0$ because patients are quickly isolated or die,
    drastically shortening the effective infectious window \citep{Althaus2014}.}
    \label{tab:example-disease}
\end{table}

\begin{intuitionbox}{Key Questions \& Takeaway}
The epidemic curve is the visible fingerprint of the hidden parameters $\beta$ and $\gamma$.
\textcolor{red}{But how do we recover those parameters from noisy outbreak data?} This is where
\textbf{Bayesian inference} comes in: rather than seeking a single best estimate, it asks
for the full distribution of parameter values that are consistent with the data and our
prior knowledge, the \emph{posterior distribution}. The practical challenge is that
this posterior cannot be computed directly, because it requires evaluating an intractable
integral over all possible parameter values.

\medskip
\textbf{Markov Chain Monte Carlo (MCMC)} is the computational engine that solves this
problem. It explores the parameter space $(\beta, \gamma)$ by running a chain of random
proposals, accepting combinations that improve the posterior probability and occasionally
accepting worse ones to avoid getting trapped at a single solution. Over many iterations, the chain maps out the full posterior distribution without ever needing to
evaluate the intractable integral directly.

\medskip
In this way, Bayesian inference provides the \emph{statistical framework} - what we want
to compute - and MCMC provides the \emph{algorithm} - how we compute it. The ODE solver
sits at the heart of both: every time MCMC proposes a new $(\beta, \gamma)$, the solver runs
the SIR model forward and compares the resulting curve to the observed case counts, scoring
how plausible that parameter combination is.
\end{intuitionbox}

\section{Bayesian Inference Setup}

Let $\theta = (\beta, \gamma)$ denote the unknown parameters (i.e, the infection and recovery parameters we wish to estimate) and $\mathbf{y}$ the observed data (daily infected counts). The Bayes' theorem states:

\begin{tcolorbox}[colback=boxblue, colframe=epidblue, halign=center]
\[
  \underbrace{P(\theta \mid \mathbf{y})}_{\text{posterior}} \;=\;
  \frac{\displaystyle\underbrace{P(\mathbf{y} \mid \theta)}_{\text{likelihood}}
        \;\cdot\;
        \underbrace{P(\theta)}_{\text{prior}}}
       {\displaystyle\underbrace{P(\mathbf{y})}_{\text{evidence}}}
\]
\end{tcolorbox}

\medskip
Each term has a concrete meaning in our epidemic setting:

\medskip
\begin{tabular}{lp{0.72\textwidth}}
\toprule
\textbf{Term} & \textbf{Meaning in the SIR context} \\
\midrule
$P(\theta)$ & Our \emph{prior} belief about $\beta$ and $\gamma$ before seeing any case data. \\[4pt]
$P(\mathbf{y} \mid \theta)$ & The \emph{likelihood}: how probable are the observed counts if the true parameters were $\theta$? \\[4pt]
$P(\theta \mid \mathbf{y})$ & The \emph{posterior}: our updated belief after seeing the outbreak data. \\[4pt]
$P(\mathbf{y})$ & A normalising constant, the same for all $\theta$, so often ignored. \\
\bottomrule
\end{tabular}

\bigskip
\begin{warningbox}{The Problem}
Computing $P(\mathbf{y}) = \int P(\mathbf{y} \mid \theta)\, P(\theta)\, d\theta$ requires integrating over \emph{all possible parameter values}. For the SIR model, this integral has no closed form. This is why we need MCMC.
\end{warningbox}

\subsection{Specifying the Likelihood}

\subsubsection{The noise model}

Real case counts are noisy: reporting delays, under-ascertainment, weekend effects. A simple and standard assumption is additive Gaussian noise

\[
  I_{\text{obs}}(t) = I_{\text{model}}(t;\,\theta) + \varepsilon_t, \qquad
  \varepsilon_t \sim \mathcal{N}(0, \sigma^2).
\]

That is, each observed count is the SIR model's prediction, plus independent Gaussian noise with standard deviation $\sigma$.

\subsubsection{Deriving the log-likelihood}

The probability of a single observation $I_{\text{obs}}(t)$ under our model is

\[
  P\!\left(I_{\text{obs}}(t) \mid \theta\right)
  = \frac{1}{\sigma\sqrt{2\pi}}
    \exp\!\left(-\frac{\left[I_{\text{obs}}(t) - I_{\text{model}}(t;\theta)\right]^2}{2\sigma^2}\right).
\]

Assuming independence across time points, the joint likelihood is the product

\[
  P(\mathbf{y} \mid \theta)
  = \prod_{t=1}^{T}
    \frac{1}{\sigma\sqrt{2\pi}}
    \exp\!\left(-\frac{r_t^2}{2\sigma^2}\right),
\]

where $r_t = I_{\text{obs}}(t) - I_{\text{model}}(t;\theta)$ is the residual at time $t$.

Taking the logarithm (and dropping constants that do not depend on $\theta$)

\begin{tcolorbox}[colback=boxblue, colframe=epidblue]
\[
  \log P(\mathbf{y} \mid \theta)
  \;\propto\;
  -\frac{1}{2} \sum_{t=1}^{T} \frac{r_t^2}{\sigma^2}
  \;=\;
  -\frac{1}{2} \sum_{t=1}^{T}
  \left(\frac{I_{\text{obs}}(t) - I_{\text{model}}(t;\theta)}{\sigma}\right)^{\!2}.
\]
\end{tcolorbox}

The log-likelihood is large (near zero) when the model fits well, and very negative when the model fits poorly.

\begin{intuitionbox}{Connection to Least Squares}
Maximising the Gaussian log-likelihood over $\theta$ is \emph{identical} to minimising the sum of squared residuals. Least squares and Gaussian likelihood are the same assumption viewed from different angles. Bayesian inference simply adds a prior on top and propagates the full uncertainty rather than reporting a single optimal value.
\end{intuitionbox}

\subsection{Specifying the Prior}

In our example we use a \textbf{uniform (flat) prior} over biologically sensible ranges

\[
  \beta \sim \text{Uniform}(0.05,\; 1.0), \qquad
  \gamma \sim \text{Uniform}(0.01,\; 0.5).
\]

These ranges encode only weak prior knowledge: we know both parameters must be positive and not astronomically large. In log-space

\[
  \log P(\theta) =
  \begin{cases}
    0       & \text{if } \theta \in \text{support}\\
    -\infty & \text{otherwise}
  \end{cases}.
\]

Returning $-\infty$ in log-space means zero probability, the MCMC sampler will never visit these regions.

\begin{warningbox}{Choosing Priors in Practice}
Flat priors are a reasonable starting point, but informative priors can be very valuable when external data exist (e.g.\ serial interval studies constrain $\gamma$, contact surveys constrain $\beta$). A well-chosen prior stabilises inference when data are sparse and naturally embeds domain knowledge.
\end{warningbox}

\section{Markov Chain Monte Carlo (MCMC)}

\subsection{The core idea}

We cannot evaluate $P(\mathbf{y})$, so we cannot compute the posterior directly. But we do not need to. We only need to \textbf{draw samples from the posterior}. Once we have samples $\theta^{(1)}, \theta^{(2)}, \ldots, \theta^{(M)}$, we can estimate any quantity of interest, e.g

\[
  \mathbb{E}[\beta \mid \mathbf{y}] \approx \frac{1}{M}\sum_{i=1}^{M} \beta^{(i)},
  \qquad
  \text{Var}[\beta \mid \mathbf{y}] \approx \frac{1}{M}\sum_{i=1}^{M}\!\left(\beta^{(i)} - \bar{\beta}\right)^{\!2}.
\]

MCMC does this by building a \textbf{Markov chain}, a sequence of parameter values where each value depends only on the previous one, whose long-run distribution is the target posterior.

\subsection{The Metropolis-Hastings Algorithm}

\begin{keybox}{Algorithm: Metropolis-Hastings}
\begin{enumerate}[nosep, leftmargin=*]
  \item \textbf{Initialise}: Choose starting values $\theta^{(0)}$.
  \item \textbf{Propose}: Draw $\theta' \sim q(\theta' \mid \theta^{(t)})$, e.g.\ a Gaussian random walk
    \[
      \beta' = \beta^{(t)} + \varepsilon_\beta, \quad
      \gamma' = \gamma^{(t)} + \varepsilon_\gamma, \quad
      \varepsilon \sim \mathcal{N}(0, \delta^2)
    \]
  \item \textbf{Accept or reject}: Compute the acceptance ratio
    \[
      \alpha = \min\!\left(1,\;
        \frac{P(\theta' \mid \mathbf{y})}{P(\theta^{(t)} \mid \mathbf{y})}
      \right)
      = \min\!\left(1,\;
        \frac{P(\mathbf{y}\mid\theta')\,P(\theta')}{P(\mathbf{y}\mid\theta^{(t)})\,P(\theta^{(t)})}
      \right)
    \]
    Note: $P(\mathbf{y})$ cancels! Set $\theta^{(t+1)} = \theta'$ with probability $\alpha$, else $\theta^{(t+1)} = \theta^{(t)}$.
  \item \textbf{Repeat} for $t = 1, 2, \ldots, N_{\text{iter}}$.
\end{enumerate}
\end{keybox}

\medskip
The key insight in Step 3 is profound: since $P(\mathbf{y})$ appears in both numerator and denominator, it \textbf{cancels out entirely}. We never need to compute the intractable normalising constant.

\subsection{Why does this work?}

\begin{itemize}
  \item If the proposed $\theta'$ has \emph{higher} posterior probability than the current $\theta^{(t)}$: always accepted ($\alpha = 1$). The chain moves uphill.
  \item If $\theta'$ has \emph{lower} posterior probability: accepted with probability $\alpha < 1$. The chain occasionally moves downhill.
\end{itemize}

This occasional downhill movement is essential, it allows the chain to \emph{explore} the full posterior shape rather than simply converging to a single peak. Over many iterations, the fraction of time the chain spends near any parameter value is proportional to that value's posterior probability.

\subsection{Burn-in}

The chain starts from an arbitrary point $\theta^{(0)}$ and takes some time to reach the high-probability region of the posterior. Samples collected during this initial ``warm-up'' period are discarded. This is called the \textbf{burn-in} period.

\[
  \underbrace{\theta^{(0)}, \theta^{(1)}, \ldots, \theta^{(B)}}_{\text{burn-in (discard)}},\;
  \underbrace{\theta^{(B+1)}, \ldots, \theta^{(N)}}_{\text{posterior samples (keep)}}
\]

\begin{intuitionbox}{Intuition for the Whole Algorithm}
Think of the posterior surface as a hilly landscape where altitude represents probability. The Metropolis-Hastings algorithm is like a hiker who always climbs uphill but occasionally stumbles downhill. Over time, the hiker spends most of their time near the peaks (high-probability regions) but occasionally visits the valleys too. The density of the hiker's footprints is a map of the posterior distribution.
\end{intuitionbox}

\section{MCMC Results and Discussion}
\subsection{Example}

We now walk through a complete implementation. We simulate synthetic outbreak data and then recover the parameters via MCMC, mimicking the real-world task of fitting a model to surveillance data.

\subsubsection{Setup}

\begin{itemize}
  \item Population: $N = 1000$, initially $I_0 = 10$ infected.
  \item True parameters: $\beta^* = 0.3$, $\gamma^* = 0.1$ $\Rightarrow$ $\mathcal{R}_0^* = 3.0$.
  \item Observed data: SIR simulation + Gaussian noise ($\sigma = 15$).
  \item Prior: $\beta \sim U(0.05, 1.0)$, $\gamma \sim U(0.01, 0.5)$.
  \item MCMC: 8000 iterations, step size $\delta = 0.015$, burn-in of 2000.
\end{itemize}

\begin{figure}[h!]
  \centering
  \includegraphics[width=\textwidth]{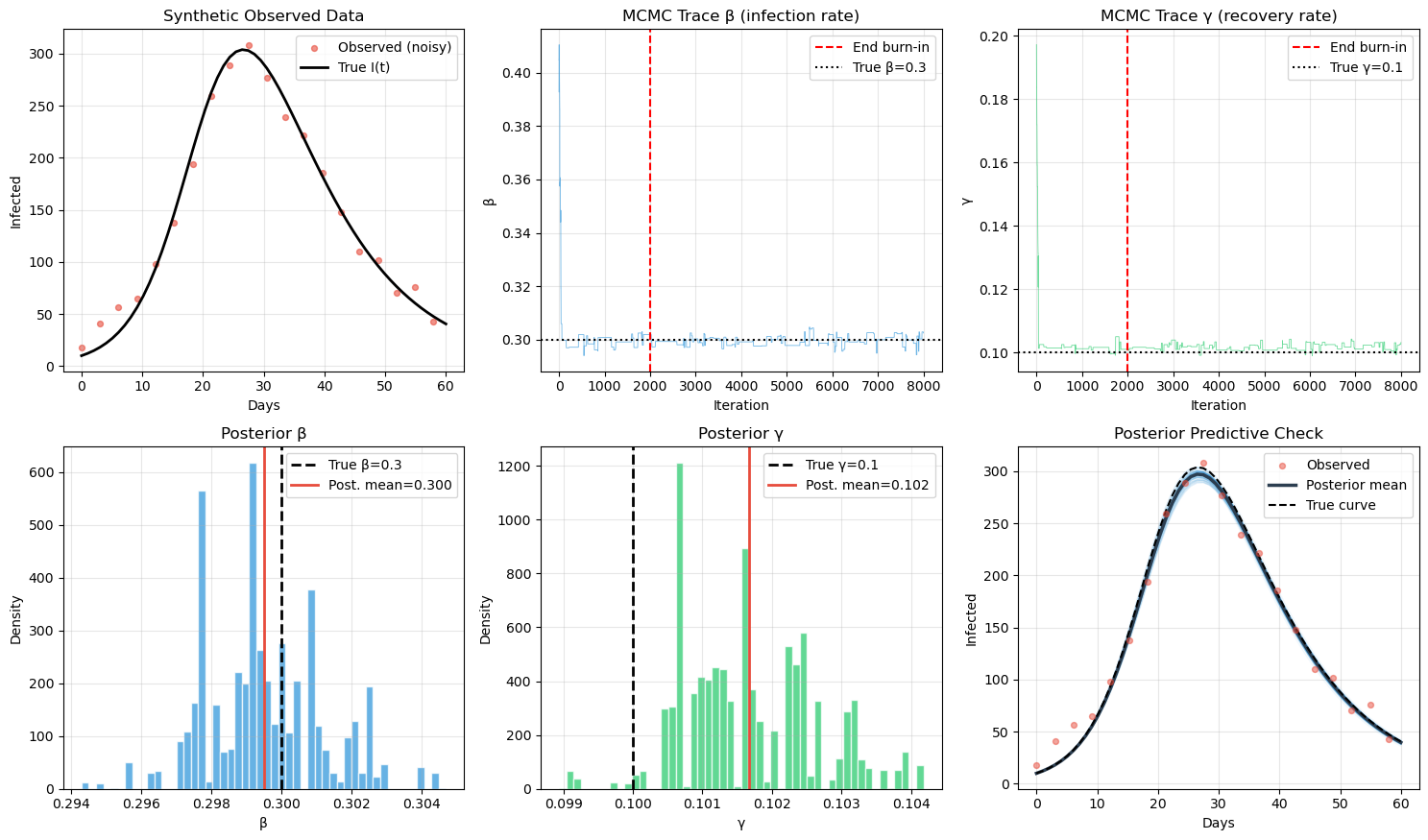}
  \caption{MCMC output for the SIR parameter estimation problem ($N=1000$,
  $\beta^*=0.3$, $\gamma^*=0.1$, $\mathcal{R}_0^*=3.0$, 8000 iterations,
  burn-in = 2000). \textit{Top row, left to right}: noisy observed infected
  counts against the true epidemic curve; MCMC trace for $\beta$; MCMC trace
  for $\gamma$. \textit{Bottom row, left to right}: posterior distribution of
  $\beta$; posterior distribution of $\gamma$; posterior predictive check
  showing 100 SIR trajectories drawn from the posterior against the observed data.}
  \label{fig:mcmc}
\end{figure}

\subsubsection{Results}

After discarding the burn-in, the posterior means recovered the true parameters accurately:

\medskip
\begin{center}
\begin{tabular}{lccc}
\toprule
\textbf{Parameter} & \textbf{True value} & \textbf{Posterior mean} & \textbf{Posterior std.} \\
\midrule
$\beta$              & 0.300 & 0.300 & 0.002 \\
$\gamma$             & 0.100 & 0.102 & 0.001 \\
$\mathcal{R}_0 = \beta/\gamma$ & 3.00 & 2.95 & 0.03 \\
\bottomrule
\end{tabular}
\end{center}
\medskip

The narrow posterior standard deviations reflect the informativeness of 60 days of daily data. More noise, fewer data points or a less identifiable model would produce wider posteriors.

\subsection{Discussing the Results}

Figure~\ref{fig:mcmc} summarises the full MCMC output. \textbf{Panel 1 (top left) - observed data:} The noisy red dots are the synthetic observed infected counts, generated by adding
Gaussian noise ($\sigma = 15$) to the true SIR trajectory. The black curve is the
ground truth. This panel contextualises the inference problem: the MCMC algorithm
sees only the noisy dots and must work backwards to recover the parameters that
plausibly generated them. The gap between the dots and the true curve represents
the irreducible observational uncertainty that the posterior will absorb.

\textbf{Panels 2 and 3 (top centre and right) - trace plots:} The trace plots show the chain's path through $\beta$ and $\gamma$ values over all
8000 iterations. A well-behaved chain resembles a dense, fast-mixing
\emph{caterpillar}, bouncing rapidly around a stable region with no long-range
drifts or flat stretches. The red dashed vertical line marks the end of the burn-in
period (iteration 2000), after which the chain is considered to have converged to
the stationary distribution. Samples before this line are discarded.

Two warning signs to watch for in practice: if the chain \emph{drifts} slowly across
the plot, it has not converged and more iterations are needed; if it gets \emph{stuck}
for long stretches at a single value, the proposal step size $\delta$ is too large and
nearly all proposals are being rejected. A healthy acceptance rate lies roughly between
20\% and 40\% for a two-parameter random walk Metropolis sampler.

\textbf{Panels 4 and 5 (bottom left and centre) - posterior distributions:} The histograms of post-burn-in samples are the central deliverable of Bayesian inference: they \emph{are} the posterior distributions of $\beta$ and $\gamma$. Several features
deserve attention. The peaks of both histograms sit very close to the true values
($\beta^*=0.3$, $\gamma^*=0.1$), confirming that the inference is accurate. The widths
of the distributions quantify posterior uncertainty: narrow peaks indicate that the data
are highly informative about the parameter, while wide peaks indicate that many parameter
values remain plausible after seeing the data. In our example the posteriors are tight
because 60 days of daily counts provide strong constraints. With noisier data, fewer
observations or a shorter outbreak window, the posteriors would broaden, and the
Bayesian framework captures this automatically, without any additional modelling effort.

\textbf{Panel 6 (bottom right) - posterior predictive check:} The posterior predictive check is the ultimate model validation tool. One hundred
$(\beta, \gamma)$ pairs are drawn at random from the posterior and each is used to
simulate a full SIR trajectory. The resulting blue band represents the full range of
epidemic curves that are consistent with both the data and our model assumptions. If
the observed data points fall comfortably within this band, the model is a credible
explanation of the outbreak. If the data systematically fall outside the band, the
model is misspecified, perhaps the SIR assumption of permanent immunity is violated
or the population is structured in ways the model ignores. The posterior predictive
check therefore bridges parameter estimation and model criticism, and should always
be reported alongside the posterior summaries \citep{Gelman2013}.

\subsection{Credible Intervals vs Confidence Intervals}

A 95\% \textbf{credible interval} $[a, b]$ means:
\[
  P(a \le \beta \le b \mid \mathbf{y}) = 0.95
\]
There is a 95\% probability, given the data and our prior, that $\beta$ lies between $a$ and $b$. This is the direct, intuitive statement most practitioners want.

A 95\% frequentist \textbf{confidence interval} does \emph{not} mean this. It means: if we repeated the experiment many times, 95\% of such intervals would contain the true value. This is a subtle and often misunderstood distinction.

In practice, for flat priors and large samples, credible intervals and confidence intervals tend to agree numerically, but their interpretations remain fundamentally different.

\subsection{Beyond Metropolis-Hastings}

The Metropolis-Hastings algorithm is the conceptual foundation, but modern epidemiological work uses more sophisticated samplers:

\medskip
\begin{tabular}{p{0.25\textwidth} p{0.65\textwidth}}
\toprule
\textbf{Sampler} & \textbf{Key idea and when to use it} \\
\midrule
\textbf{NUTS / HMC} & Uses gradient information to propose distant moves efficiently. The default in Stan and PyMC. Much better for higher-dimensional problems. \\[6pt]
\textbf{Gibbs sampling} & Updates one parameter at a time from its conditional distribution. Useful when conditionals have closed forms. \\[6pt]
\textbf{SMC} & Sequential Monte Carlo, well-suited to time-series and online updating as data arrives. \\[6pt]
\textbf{NUTS in Stan} & The practical recommendation for most SIR/SEIR fitting problems. Very few tuning parameters, excellent diagnostics. \\
\bottomrule
\end{tabular}

\subsection{Summary and Key Takeaways}

\begin{keybox}{The Five Things to Remember}
\begin{enumerate}[leftmargin=*, nosep]
  \item \textbf{Bayesian inference is about distributions, not point estimates.} The posterior $P(\theta \mid \mathbf{y})$ tells you the full range of plausible parameter values, not just a single best guess.
  \item \textbf{The three ingredients are prior, likelihood, and Bayes' theorem.} The prior encodes pre-existing knowledge; the likelihood scores how well each parameter value explains the data; their product (normalised) is the posterior.
  \item \textbf{The Gaussian log-likelihood is just penalised squared residuals.} If you have used least squares before, you have implicitly assumed Gaussian noise, Bayesian inference makes this explicit.
  \item \textbf{MCMC avoids the intractable normalising constant.} The Metropolis-Hastings acceptance ratio cancels $P(\mathbf{y})$, so we can sample from the posterior without ever computing it.
  \item \textbf{Always check your trace plots and do a posterior predictive check.} Trace plots diagnose sampler health; posterior predictive checks verify that the fitted model actually reproduces the observed data.
\end{enumerate}
\end{keybox}

\bigskip
\begin{intuitionbox}{The Big Picture}
Epidemiological models are tools for understanding transmission dynamics and informing policy. Their parameters are never known exactly. Bayesian MCMC gives us a principled, computationally feasible way to say not just \emph{``our best estimate of $\mathcal{R}_0$ is 2.5''} but \emph{``given the data, $\mathcal{R}_0$ is between 2.1 and 2.9 with 95\% probability''}, and that uncertainty matters enormously for intervention planning.
\end{intuitionbox}

\newpage
\bibliographystyle{plainnat}

\end{document}